\documentclass[pdflatex,sn-vancouver-num]{sn-jnl}
\usepackage{graphicx}
\graphicspath{{./figures/}}
\usepackage{amsmath,amssymb,amsfonts}
\usepackage{mathtools}
\usepackage{amsthm}
\usepackage[title]{appendix}
\usepackage{tabularx}
\usepackage{booktabs}
\usepackage{multirow}
\usepackage{placeins}
\usepackage{float}
\usepackage{cleveref}

\DeclareMathOperator{\logit}{logit}
\theoremstyle{thmstylethree}
\newtheorem{example}{Example}[section]

\usepackage{setspace}
\usepackage{geometry}
\begin{document}

\title[Generalized Odds for Ordinal Outcomes]{Generalized Odds for Ordinal Outcomes: A Causal Inference Framework Adjusting for Covariates}

\author[1]{\fnm{Runnian} \sur{Huang}}
\author[2]{\fnm{Dandan} \sur{Shi}}
\author*[1,2]{\fnm{Lan} \sur{Liu}}\email{liulan1815@outlook.com}
\author*[1,3]{\fnm{Yunjian} \sur{Yin}}\email{yinyj0603@163.com}

\affil[1]{\orgname{Beijing Institute of Brain Disorders, Capital Medical University}, \orgaddress{\city{Beijing}, \postcode{100069}, \country{China}}}
\affil[2]{\orgname{Chinese Institutes for Medical Research}, \orgaddress{\city{Beijing}, \postcode{100070}, \country{China}}}
\affil[3]{\orgdiv{Department of Neurology}, \orgname{Xuanwu Hospital, Capital Medical University}, \orgaddress{\city{Beijing}, \postcode{100053}, \country{China}}}

\abstract{\textbf{Background:}
Generalized Odds (Gen Odds) and related tournament methods provide useful summaries for ordinal outcomes without requiring the proportional odds assumption. However, conventional Gen Odds is formulated as a marginal comparison between independently drawn treatment and control outcomes and therefore targets a population-level contrast rather than an individual-level contrast based on the joint distribution of potential outcomes. We aimed to develop an individual-level causal formulation of Gen Odds and establish conditions for its identification.

\textbf{Methods:}
We defined the Causal Gen Odds using the joint distribution of potential outcomes and established its identification under conditional independence of the potential outcomes given baseline covariates. We developed a model-based estimator using nonproportional cumulative logistic regression and introduced an odds-ratio sensitivity parameter to assess departures from conditional independence. The proposed estimators were evaluated through simulation studies and illustrated using data from the ACCORD Study.

\textbf{Results:}
In simulations, the conditional-independence estimator was approximately unbiased with 95.5\% confidence interval coverage when conditional independence held, but bias and undercoverage increased as residual dependence departed from independence. When the sensitivity-analysis estimator incorporated the data-generating dependence parameter, bias remained below 0.01 and coverage ranged from 93.7\% to 96.1\% across all examined settings. In the ACCORD Study, the Causal Gen Odds for right-foot pre-tibial edema was 0.909 (95\% CI: 0.880 to 0.940) under conditional independence and remained stable across the examined sensitivity range.

\textbf{Conclusions:}
The Causal Gen Odds provides an individual-level estimand that complements the conventional population-level Gen Odds. The proposed identification and sensitivity-analysis framework enables estimation of this contrast while explicitly assessing its robustness to assumptions about dependence between potential outcomes.}


\maketitle

\section{Background}
Randomized controlled trials (RCTs) frequently use ordinal scales, such as the Expanded Disability Status Scale (EDSS) for multiple sclerosis and the Modified Rankin Scale (mRS) for stroke, to assess clinical outcomes \cite{kurtzke1983rating,van1988interobserver}. The standard approach for analyzing such data is the proportional odds logistic regression (POLR) model. This model relies on the proportional odds assumption, which requires that the treatment effect—typically expressed as a common odds ratio (cOR)—remains constant across all possible thresholds of the ordinal scale \cite{brant1990assessing}. In clinical practice, however, this assumption is frequently violated, as treatment effects often differ across thresholds of the scale. For instance, a therapy evaluated via the mRS might demonstrate a robust effect on achieving functional independence (mRS 0–2 vs. 3–6), but a comparatively modest impact on preventing severe disability or death (mRS 0–4 vs. 5–6). Although some researchers attempt to circumvent this issue by dichotomizing ordinal scores into binary outcomes, such practice generally results in substantial loss of statistical power and information \cite{armstrong1989ordinal,tutz2011regression,selman2023statistical,d2020ordinal}. These limitations have motivated alternative methods that preserve the ordinal structure of the data.

The Cochran–Mantel–Haenszel (CMH) test is a generalized rank-based test for assessing a shift in the ordinal outcome distribution without assuming proportional odds; its extension, the van Elteren test, accommodates stratification by baseline variables \cite{lees2006nxy,koch2000categorical}. However, the test yields only a p-value without providing an estimate of the treatment effect. Howard et al.\ proposed a permutation-based approach for analyzing mRS that yields a robust p-value but does not inherently provide a confidence interval (CI) \cite{howard2012simple}. Methods such as Agresti generalized Odds (Gen Odds) and Wilcoxon-Mann-Whitney (WMW) Gen Odds (hereafter, conventional Gen Odds) overcome these limitations: without invoking the proportional odds assumption, they provide both an interpretable effect size and its associated CIs, thereby facilitating a more detailed evaluation of clinical efficacy \cite{agresti1980generalized,o2006exploiting,churilov2014improved,howard2022wilcoxon}. Within this family, the Agresti Gen Odds excludes tied pairs from the calculation, which can substantially reduce the effective sample size and yield less stable estimates in ordinal outcome settings. We therefore adopt the WMW Gen Odds formulation in this study, which handles ties by equal splitting, avoids sample-size reduction, and yields more conservative, robust estimates \cite{churilov2014improved,howard2022wilcoxon}. These methods, collectively referred to as ``Tournament Methods'', perform all-to-all pairwise comparisons between treatment and control patients \cite{churilov2022tournament}; the family also includes the Win Ratio, Win Difference \cite{gregson2025use}, Win Probability (WinP) \cite{zou2022distribution}, and a method based on the Mann-Whitney statistic \cite{koziol2006analysis}.

Despite their utility, ``Tournament methods" must be carefully interpreted when their estimates are given a causal meaning. By construction, they compare the marginal outcome distributions of two independent groups, pairing an outcome from the treatment group with an independently drawn outcome from the control group. This comparison answers a well-defined population-level question: does the treatment shift the outcome distribution in the target population? This is analogous to the average treatment effect (ATE), which compares the mean outcome under treatment versus control without requiring knowledge of how these two potential outcomes co-vary at the individual level.

Although the marginal comparison is a population-level summary, the win probability—the chance that a randomly selected treated patient has a better outcome than a randomly selected control patient—is sometimes interpreted as an individual-level probability of benefit. Answering this individual-level question requires knowledge of how the two potential outcomes co-vary within the same patient—that is, their joint distribution. The conventional marginal comparison does not use this joint information; it instead contrasts two randomly selected individuals from the respective treatment groups. In clinical settings, potential outcomes are typically positively correlated through shared prognostic factors; when this dependence is strong, the marginal comparison can produce estimates that differ substantially from the individual-level contrast—and in some cases, may even be in the opposite direction \cite{rubin1974estimating,holland1986statistics}.

Existing covariate-adjustment methods do not bridge this gap. The stratified win ratio compares outcomes within covariate-homogeneous strata and pools stratum-specific estimates, thereby controlling for measured confounders \cite{dong2018stratified, gasparyan2021adjusted}; inverse probability weighting reweights subjects to a pseudo-population with balanced baseline covariates \cite{wang2025adjusted}. While these approaches target causal effects at the population level—the average effect of treatment on the outcome distribution—they do not provide information about how the two potential outcomes co-vary within individuals. Consequently, they address a different question than the individual-level contrast—the probability that a patient has a better outcome under treatment than under control. Neither uses the joint distribution required for the latter.

Building on these considerations, this study first formally clarifies the distinction between the population-level estimand targeted by conventional Gen Odds and the individual-level estimand defined through the joint distribution of potential outcomes. We then develop a causal estimand for the Gen Odds that targets the individual-level contrast—the probability that a patient benefits from treatment—under the joint distribution of potential outcomes, complementing the population-level summary provided by conventional methods. Finally, we introduce a sensitivity analysis framework employing a global odds ratio to quantify the impact of varying degrees of potential outcome dependence, thereby evaluating the robustness of causal conclusions to uncertainty about the unobserved joint distribution.

\subsection{Motivating Examples}
	
The following two examples illustrate two distinct ways in which population-level marginal comparisons can differ from individual-level contrasts when the potential outcomes are dependent: the apparent treatment effect may be reversed in direction or substantially attenuated toward the null.

\begin{example} \label{Example 1}
Let $Y_c$ and $Y_t$ denote the potential outcomes under control and treatment, respectively. Consider a hypothetical scenario where the control potential outcome follows a standard normal distribution,
\[
Y_c \sim N(0, 1).
\]
The treatment potential outcome is a discontinuous function of the control outcome, defined as
\[
Y_t = 
\begin{cases} 
Y_c + 5, & Y_c < -0.5, \\
Y_c - 0.5, & Y_c \geq -0.5.
\end{cases}
\]
This construction implies that for patients with very poor control outcomes ($Y_c < -0.5$), the treatment provides a large benefit of $+5$; for patients with better control outcomes ($Y_c \geq -0.5$), the treatment is mildly harmful, reducing the outcome by $0.5$.

Under the joint dependence structure, the individual-level win probability is computed as
\[
\Pr(Y_t > Y_c) = 31\%.
\]
That is, although the treatment helps the worst-off subgroup, it hurts the majority, resulting in a net probability that the treatment beats control of less than half.

However, the conventional ``Tournament method" treats the two potential outcomes as independent draws from their respective marginal distributions, thereby addressing a different question than the individual-level contrast. Because the marginal comparison is constructed from independent draws, it does not preserve the deterministic dependence between \(Y_t\) and \(Y_c\) imposed by the data-generating rule above. Under this marginal comparison, the population-level summary yields
\[
\Pr(\tilde{Y}_t > \tilde{Y}_c) = 65\%,
\]
where $\tilde{Y}_t$ and $\tilde{Y}_c$ are mutually independent draws from the marginal distributions of $Y_t$ and $Y_c$, respectively.

Thus, the joint comparison yields a 31\% win probability, which addresses the individual-level question of whether a randomly selected patient benefits from treatment. The marginal comparison, in contrast, yields 65\%, which addresses a different population-level question: whether a randomly selected treated patient has a better outcome than a randomly selected control patient. Notably, the marginal comparison suggests that the treatment is effective (win probability $>$ 50\%; conventional Gen Odds $=1.86>1$), whereas the joint comparison—the relevant benchmark for individual benefit—indicates that the treatment is actually harmful for the majority of patients (win probability $<$ 50\%; Causal Gen Odds $=0.45<1$). This reversal demonstrates that evaluating treatment efficacy solely based on marginal effects is insufficient; a treatment that appears beneficial at the population level may be ineffective or even harmful at the individual level.
\end{example}

\begin{example}
\label{example_heterogeneity}
Whereas Example \ref{Example 1} illustrates a reversal in the direction of the
treatment effect, a different discrepancy can arise when substantial
prognostic heterogeneity attenuates the marginal population-level
comparison toward the null, even when a substantial individual-level
treatment effect is present. Let
\[
X \sim N(0,0.5^2),
\]
and suppose that the potential outcomes are generated as
\[
Y_c = 10X+\varepsilon_c,
\qquad
Y_t = 10X+\delta+\varepsilon_t,
\]
where $\delta=0.1$ is the treatment effect and
$\varepsilon_c,\varepsilon_t \overset{\mathrm{iid}}{\sim}N(0,0.1^2)$
are independent residual terms.

For the same individual, the common prognostic component $10X$
cancels from the treatment contrast:
\[
Y_t-Y_c
=
\delta+\varepsilon_t-\varepsilon_c
\sim N(0.1,0.02).
\]
Thus, the individual-level probability that treatment produces a
more favorable outcome than control is
\[
\Pr(Y_t>Y_c)
=
\Phi\left(\frac{0.1}{\sqrt{0.02}}\right)
=
\Phi(0.707)
\approx 76.0\%.
\]
For a continuous outcome, where ties occur with probability zero,
this corresponds to a Causal Gen Odds of
\[
\frac{0.760}{1-0.760}\approx 3.17.
\]

In contrast, the conventional marginal comparison considers
independent draws from the treatment and control outcome
distributions. The prognostic components of the two independently
drawn individuals no longer cancel. Since
\[
\operatorname{Var}(10X)=25,
\]
the variance of the marginal pairwise contrast becomes
\[
2\operatorname{Var}(10X)
+\operatorname{Var}(\varepsilon_t)
+\operatorname{Var}(\varepsilon_c)
=
50.02.
\]
Consequently,
\[
\Pr(\widetilde{Y}_t>\widetilde{Y}_c)
=
\Phi\left(\frac{0.1}{\sqrt{50.02}}\right)
\approx 50.6\%,
\]
corresponding to a conventional Gen Odds of approximately 1.02.

Hence, although the marginal population-level comparison is nearly null, the individual-level comparison indicates a substantial treatment benefit. This example illustrates that a near-null marginal Gen Odds does not necessarily imply the absence of an individual-level treatment effect; strong prognostic heterogeneity can make the two estimands differ substantially. Together, Examples~\ref{Example 1} and~\ref{example_heterogeneity} show that marginal and individual-level comparisons may differ not only in magnitude but also in direction. This motivates a formal distinction between the two corresponding Gen Odds estimands, which we introduce in the next section.
\end{example}

\section{Methods}

\subsection{Generalized Odds: Notation and Estimands}

Let $T$ denote the binary treatment assignment, where $T=t$ indicates the treatment group and $T=c$ the control group, and let $Y$ be the ordinal outcome with levels $1,\ldots,J$. Without loss of generality, we assume that larger values indicate better clinical outcomes; for outcomes coded in the opposite direction, the interpretation of the Gen Odds is reversed. Let $Y_t$ and $Y_c$ denote the potential outcomes that would be observed if the individual received treatment or control, respectively \cite{rubin1974estimating}. The conventional Gen Odds is computed solely from the marginal distributions of $Y_t$ and $Y_c$ by comparing independent draws from these two distributions:

\begin{equation} \label{eq1}
\textnormal{Conventional Gen Odds} = \frac{\Pr(\tilde{Y}_t > \tilde{Y}_c) + 
\Pr(\tilde{Y}_t = \tilde{Y}_c)/2}{\Pr(\tilde{Y}_t < \tilde{Y}_c) 
+ \Pr(\tilde{Y}_t = \tilde{Y}_c)/2},
\end{equation}
where $(\tilde{Y}_t, \tilde{Y}_c)$ are independent draws from the respective marginal distributions of $Y_t$ and $Y_c$. This quantity summarizes the population-level difference between the two marginal outcome distributions. It can be interpreted as the odds that an independently drawn treated outcome is more favorable than an independently drawn control outcome, relative to the reverse comparison, with ties split equally between the two directions. However, it does not address whether the same individual would have a more favorable outcome under treatment than under control, because it does not involve the joint distribution of $(Y_t, Y_c)$ \cite{rubin1974estimating,holland1986statistics}.

To develop a causal estimand, we adopt the potential outcomes framework \cite{rubin1974estimating,holland1986statistics}. We retain the algebraic form of Equation~(\ref{eq1}) but replace the independent copies $(\tilde{Y}_t, \tilde{Y}_c)$ with the potential outcome pair $(Y_t, Y_c)$ for the same randomly selected individual:

\begin{equation} \label{eq2}
\textnormal{Causal Gen Odds} = \frac{\Pr(Y_{t} > Y_{c}) + \Pr(Y_{t} = Y_{c})/2}
{\Pr(Y_{t} < Y_{c}) + \Pr(Y_{t} = Y_{c})/2}.
\end{equation}

The Causal Gen Odds compares the odds of a treatment-favorable individual-level contrast with the odds of a control-favorable individual-level contrast, with ties split equally between the two directions. Specifically, values greater than 1 indicate that treatment-favorable outcomes are more likely than control-favorable outcomes for the same randomly selected individual, whereas values less than 1 indicate the reverse.

The critical distinction between Equations~(\ref{eq1}) and~(\ref{eq2}) lies in the joint distribution $q_{ij} = \Pr(Y_t = i,\,Y_c = j)$. For an ordinal outcome $Y \in \{1,2,3\}$ representing ``unchanged'', ``improved'', and ``cured'', Table~\ref{tab:1} displays this joint distribution, with treatment wins in the lower-left ($q_{21}, q_{31}, q_{32}$), treatment losses in the upper-right ($q_{12}, q_{13}, q_{23}$), and ties on the diagonal ($q_{11}, q_{22}, q_{33}$). For the conventional Gen Odds, the comparison probabilities are formed from the product of the two marginal distributions, $\Pr(\tilde{Y}_t=i,\tilde{Y}_c=j) =\Pr(Y_t=i)\Pr(Y_c=j)$. By contrast, the Causal Gen Odds depends on the actual within-individual joint probabilities $q_{ij}=\Pr(Y_t=i,Y_c=j)$, which need not factorize.

The components of Equation~(\ref{eq2}) are:
\begin{align*}
\Pr(Y_{t} > Y_{c}) &= q_{21}+q_{31}+q_{32}, \\
\Pr(Y_{t} < Y_{c}) &= q_{12}+q_{13}+q_{23}, \\
\Pr(Y_{t} = Y_{c}) &= q_{11}+q_{22}+q_{33}.
\end{align*}

\begin{table*}[!htbp]
	\centering
	\caption{Joint distribution of the potential outcomes under treatment and control}
	\label{tab:1}
	\begin{tabularx}{\textwidth}{>{\centering\arraybackslash}X>{\centering\arraybackslash}X>{\centering\arraybackslash}X>{\centering\arraybackslash}X}
		\toprule
		& \multicolumn{3}{c}{Control group ($j$)} \\
		\midrule
		Treatment group ($i$) & 1 & 2 & 3 \\
		1 & $q_{11}$ & $q_{12}$ & $q_{13}$ \\
		2 & $q_{21}$ & $q_{22}$ & $q_{23}$ \\
		3 & $q_{31}$ & $q_{32}$ & $q_{33}$ \\
		\bottomrule
	\end{tabularx}%
\end{table*}% 

The distinction between the two estimands becomes practically important when \(Y_t\) and \(Y_c\) are dependent. Because the conventional Gen Odds is determined solely by the marginal distributions whereas the Causal Gen Odds depends on their joint distribution, the two measures need not agree and may even lead to opposite conclusions. This distinction raises an identification problem: \((Y_t,Y_c)\) is never jointly observed for the same individual, even in a perfectly conducted RCT \cite{holland1986statistics}: each patient receives either treatment or control, never both. Consequently, the \(q_{ij}\) are not directly identifiable from observed data, and computing the Causal Gen Odds via Equation~(\ref{eq2}) requires an additional assumption on the joint distribution of \((Y_t,Y_c)\). The conventional Gen Odds does not require identification of \(q_{ij}\), because it is defined directly through independent draws from the two marginal distributions. The corresponding comparison probabilities therefore take the product form \(\Pr(\tilde{Y}_t=i,\tilde{Y}_c=j)=\Pr(Y_t=i)\Pr(Y_c=j)\). This yields a well-defined population-level summary without imposing independence on the actual within-individual potential outcomes. However, as illustrated in Example~\ref{Example 1}, this population-level summary can differ markedly from the individual-level contrast \(\Pr(Y_t > Y_c)\) when the potential outcomes are strongly dependent. The two quantities answer different questions; conflating them can lead to misleading conclusions about individual treatment benefit. The following sections develop a framework for identifying and estimating the Causal Gen Odds under assumptions about this joint distribution.

\subsection{Causal Identification under Conditional Independence}

As noted in Section~2.1, the joint distribution of $(Y_t, Y_c)$ is not directly identifiable from observed data. A principled route to identification is to assume that there exists a vector of baseline covariates $X$ rich enough to capture the joint dependence between the two potential outcomes, so that they are conditionally independent given $X$ \cite{shen2013treatment,zhang2013assessing}:
\begin{equation*}
	Y_t \perp\!\!\!\perp Y_c \mid X.
\end{equation*}

Under this assumption, the conditional joint distribution factorizes:
\begin{equation}
	\Pr(Y_t = i,\, Y_c = j \mid X)
	= \Pr(Y_t = i \mid X)\,\Pr(Y_c = j \mid X).
	\label{eq:factorize}
\end{equation}

Taking expectation over $X$ and then substituting Equation~\eqref{eq:factorize} yields
\begin{align}
	\Pr(Y_t > Y_c)
	&= \sum_{i>j}\Pr(Y_t=i,\,Y_c=j)
	\notag\\
	&= \sum_{i>j}\mathbb{E}\bigl[\Pr(Y_t=i,Y_c=j\mid X)\bigr]
	\notag\\
	&= \sum_{i>j}\mathbb{E}\bigl[\Pr(Y_t=i\mid X)\Pr(Y_c=j\mid X)\bigr].
	\label{eq3}
\end{align}

Under randomization and the consistency assumption ($Y = Y_T$, i.e., the observed outcome equals the potential outcome under the assigned treatment), we have
\begin{align}
	\Pr(Y_t=i\mid X) &= \Pr(Y=i\mid T=t,\,X), \notag\\
	\Pr(Y_c=j\mid X) &= \Pr(Y=j\mid T=c,\,X). \notag
\end{align}
Substituting into Equation~\eqref{eq3} gives
\begin{align}
    \Pr(Y_t > Y_c)
    =
    \sum_{i>j}
    \int
    \Pr(Y=i\mid T=t,X=x)\,
    \Pr(Y=j\mid T=c,X=x)\,
    f_X(x)\,dx,
    \label{eq:identification}
\end{align}
where $f_X(x)$ denotes the marginal distribution of $X$, encompassing both discrete and continuous cases.

Similarly, $\Pr(Y_t < Y_c)$ and $\Pr(Y_t = Y_c)$ can be derived, and the Causal Gen Odds is obtained by substituting these probabilities into Equation~\eqref{eq2}.

\begin{example}[Reversal between Conventional and Causal Gen Odds]
\label{example:reversal}

The following two cases illustrate how the conventional and Causal Gen Odds can lead to opposite conclusions and how the Causal Gen Odds can be identified under the conditional independence assumption. In both cases, we consider a binary covariate \(X\) with equal stratum probabilities, \(\Pr(X=x_1) = \Pr(X=x_2) = 0.5\) (detailed calculations are provided in \autoref{appendix:calc}).

\textbf{Case 1: Population-Level Harm, Individual-Level Benefit.}
\autoref{tab:5} presents the joint distribution of the potential outcomes in the overall population (upper panel) and within each covariate stratum (lower panels). The conventional Gen Odds, computed from independent draws from the two marginal distributions, is \(0.90\), suggesting a detrimental treatment at the population level. In contrast, evaluating the within-individual joint distribution via Equation~\eqref{eq2} gives a Causal Gen Odds of \(1.13\), indicating a beneficial treatment at the individual level. Thus, the two estimands lead to opposite conclusions regarding treatment benefit. Within each level of \(X\), \(Y_t\) and \(Y_c\) are conditionally independent, so the joint distribution factorizes according to Equation~\eqref{eq:factorize}. Applying Equation~\eqref{eq:identification} and averaging over the two equally weighted covariate strata therefore identifies the Causal Gen Odds of \(1.13\).

\begin{table*}[!htbp]
	\centering
	\caption{Case 1: under the conditional independence assumption.}
	\label{tab:5}
	\begin{tabularx}{\textwidth}{XXXXX}
		\toprule
		Subgroup & &\multicolumn{3}{c}{Control ($j$)} \\
		\midrule
		Total &Treatment ($i$) & 1 & 2 & 3 \\ 
		&1 & 0.1200 & 0.0276 & 0.2424 \\
		&2 & 0.0400 & 0.0360 & 0.0740 \\
		&3 & 0.0800 & 0.2864 & 0.0936 \\
		& & & & \\
		Covariate levels & &\multicolumn{3}{c}{Control ($j$)} \\
		X = $x_{1}$ &Treatment ($i$) & 1 & 2 & 3 \\ 
		(Pr = 0.5) & 1 & 0.2304 & 0.0144 & 0.4752 \\
		& 2 & 0.0640 & 0.0040 & 0.1320 \\
		& 3 & 0.0256 & 0.0016 & 0.0528 \\
		& &\multicolumn{3}{c}{Control ($j$)} \\
		X = $x_{2}$ &Treatment ($i$) & 1 & 2 & 3 \\ 
		(Pr = 0.5) & 1 & 0.0096 & 0.0408 & 0.0096 \\
		& 2 & 0.0160 & 0.0680 & 0.0160 \\
		& 3 & 0.1344 & 0.5712 & 0.1344 \\
		\bottomrule
	\end{tabularx}%
\end{table*}%

\textbf{Case 2: Population-Level Benefit, Individual-Level Harm.}
\autoref{tab:6} presents the symmetric reversal. The conventional Gen Odds based on the two marginal distributions is \(1.12\), suggesting a beneficial treatment at the population level, whereas the Causal Gen Odds computed from the within-individual joint distribution is \(0.85\), indicating a harmful treatment at the individual level. Again, \(Y_t\) and \(Y_c\) are conditionally independent within each level of \(X\), and applying Equation~\eqref{eq:identification} identifies the Causal Gen Odds of \(0.85\). Together, these two cases illustrate that the conventional and Causal Gen Odds answer different questions and can lead to opposite conclusions.

\begin{table*}[!htbp]
	\centering
	\caption{Case 2: under the conditional independence assumption.}
	\label{tab:6}
	\begin{tabularx}{\textwidth}{XXXXX}
		\toprule
		Subgroup & &\multicolumn{3}{c}{Control ($j$)} \\
		\midrule
		Total &Treatment ($i$) & 1 & 2 & 3 \\ 
		&1 & 0.0084 & 0.2796 & 0.0120 \\
		&2 & 0.0086 & 0.1656 & 0.0158 \\
		&3 & 0.1230 & 0.0948 & 0.2922 \\
		& & & & \\
		Covariate levels & &\multicolumn{3}{c}{Control ($j$)} \\
		X = $x_{1}$ &Treatment ($i$) & 1 & 2 & 3 \\ 
		(Pr = 0.5) & 1 & 0.0052 & 0.0024 & 0.0124 \\
		& 2 & 0.0104 & 0.0048 & 0.0248 \\
		& 3 & 0.2444 & 0.1128 & 0.5828 \\
		& &\multicolumn{3}{c}{Control ($j$)} \\
		X = $x_{2}$ &Treatment ($i$) & 1 & 2 & 3 \\ 
		(Pr = 0.5) & 1 & 0.0116 & 0.5568 & 0.0116 \\
		& 2 & 0.0068 & 0.3264 & 0.0068 \\
		& 3 & 0.0016 & 0.0768 & 0.0016 \\
		\bottomrule
	\end{tabularx}%
\end{table*}%

\end{example}

The conditional probabilities \(\Pr(Y=j\mid T,X)\) required for the identification result in Equation~\eqref{eq:identification} are typically unknown; Section~2.3 details how they are estimated from observed data.

\subsection{Model-based Estimation}

To estimate the conditional category probabilities $\Pr(Y = j \mid T, X)$ required in Section~2.2, we adopt a cumulative probability model that relaxes the proportional odds assumption. For an ordinal outcome with $J$ levels labeled $1, 2, \ldots, J$, the model specifies, for each cut-point $j = 1, 2, \ldots, J-1$,
\begin{equation}
    \mathrm{logit}\bigl\{\Pr(Y \leq j \mid T, X)\bigr\}
    = \alpha_j - \beta_{T,j}\, T - \sum_k\beta_{k,j}\, X_k,
    \label{eq:model}
\end{equation}
where $\alpha_j$ is a cut-point--specific intercept, and $\beta_{T,j}$, $\beta_{k,j}$ are cut-point--specific coefficients for treatment and covariate, respectively. Unlike the proportional odds model, the coefficients $(\beta_{T,j}, \beta_{k,j})$ are permitted to vary across $j$, relaxing the parallel-lines assumption. Under this parameterization, $\beta_{T,j} > 0$ corresponds to treatment shifting the distribution of $Y$ toward higher categories.

In practice, Equation~\eqref{eq:model} amounts to fitting $J-1$ separate binary logistic regressions, one for each dichotomization $Y \leq j$ versus $Y > j$. The per-category probabilities are then recovered by differencing:
\begin{equation*}
    \Pr(Y = j \mid T, X)
    = \Pr(Y \leq j \mid T, X) - \Pr(Y \leq j-1 \mid T, X),
    \qquad j = 1, \ldots, J,
\end{equation*}
with the boundary conventions $\Pr(Y \leq 0 \mid T, X) = 0$ and $\Pr(Y \leq J \mid T, X) = 1$. We note that when $X$ comprises only categorical variables with a manageable number of strata and sufficient observations per cell, the conditional probabilities $\Pr(Y \le j \mid T, X)$ can be estimated empirically as the within-stratum proportion of subjects with $Y \le j$ in each treatment arm, without requiring the parametric model in Equation~\eqref{eq:model}. In this case the proposed plug-in estimator reduces to a stratified nonparametric estimator that requires no modeling assumptions on the marginal distributions. The cumulative logit model in Equation~\eqref{eq:model} is required when $X$ includes continuous variables, or when $X$ has many categorical strata that produce sparse cells.

The expectation in Equation~\eqref{eq:identification} is estimated nonparametrically by the empirical distribution of $X$. Substituting the fitted probabilities yields the plug-in estimator
\begin{equation}
    \widehat{\Pr}(Y_t > Y_c)
    = \frac{1}{n}\sum_{l=1}^{n} \sum_{i > j}
        \widehat{\Pr}(Y = i \mid T = t,\, X = x_l)\,
        \widehat{\Pr}(Y = j \mid T = c,\, X = x_l),
    \label{eq:estimator}
\end{equation}
where $x_1, \ldots, x_n$ are the observed covariate values and $\widehat{\Pr}(Y = j \mid T, X)$ is obtained from the fitted model via Equation~\eqref{eq:model}. The estimators $\widehat{\Pr}(Y_t < Y_c)$ and $\widehat{\Pr}(Y_t = Y_c)$ are computed analogously, and the Causal Gen Odds estimate follows by substitution into Equation~\eqref{eq2}, giving
\[
  \widehat{\textnormal{Gen Odds}}
  = \frac{\widehat{\Pr}(Y_t > Y_c) + \widehat{\Pr}(Y_t = Y_c)/2}
         {\widehat{\Pr}(Y_t < Y_c) + \widehat{\Pr}(Y_t = Y_c)/2}.
\] The estimator developed above relies on the conditional independence assumption of Section~2.2; the next subsection introduces an estimation approach that accommodates dependence between the potential outcomes through a sensitivity parameter.

\subsection{Sensitivity Analysis for Departures from Conditional Independence}
	
The causal identification framework in Section~2.2 assumes that conditioning on $X$ removes all residual dependence between the potential outcomes $Y_t$ and $Y_c$. In practice, identifying a fully sufficient covariate set is difficult, and unmeasured factors may still link $Y_t$ and $Y_c$ after adjustment \cite{yin2018assessing}. Because residual dependence cannot be verified from observed data alone, a sensitivity analysis is needed to assess how robust the Causal Gen Odds estimate is to departures from the conditional independence assumption.

To quantify residual dependence, we introduce a sensitivity parameter $\gamma_{i,j,x}$, defined as the odds ratio of the $2 \times 2$ table obtained by dichotomizing the potential outcomes at cut-points $(i, j)$ within covariate stratum $X = x$ (see \autoref{tab:7}) \cite{shen2013treatment,williamson1995analyzing}:
\begin{align}
	\gamma_{i, j, x} &= \frac{\Pr(Y_{t} < i, Y_{c} < j \mid x) \Pr(Y_{t} \ge i, Y_{c} \ge j \mid x)}{\Pr(Y_{t} \ge i, Y_{c} < j \mid x) \Pr(Y_{t} < i, Y_{c} \ge j \mid x)}.\label{eq4}
\end{align}

\begin{table*}[!htbp]
	\centering
	\caption{The joint distribution of ordinal outcomes.}
	\label{tab:7}
	\begin{tabularx}{\textwidth}{lXXXc}
		\toprule
		& & $Y_{t} < i$ & $Y_{t} \ge i$ & Total \\
		\midrule
		& $Y_{c} < j$ & $\Pr(Y_{t} < i, Y_{c} < j \mid x)$ & $\Pr(Y_{t} \ge i, Y_{c} < j \mid x)$ & $\Pr(Y_{c} < j \mid x)$ \\
		& $Y_{c} \ge j$ & $\Pr(Y_{t} < i, Y_{c} \ge j \mid x)$ & $\Pr(Y_{t} \ge i, Y_{c} \ge j \mid x)$ & $1 - \Pr(Y_{c} < j \mid x)$ \\
		& Total & $\Pr(Y_{t} < i \mid x)$ & $1 - \Pr(Y_{t} < i \mid x)$ & 1 \\
		\bottomrule
	\end{tabularx}%
\end{table*}%

The case $\gamma_{i,j,x} = 1$ for all $(i, j, x)$ corresponds to conditional independence and recovers the estimator of Section~2.2. A value $\gamma_{i,j,x} > 1$ indicates positive residual association at the dichotomization $(i, j)$: the events $\{Y_t < i\}$ and $\{Y_c < j\}$ co-occur more often than expected under independence, reflecting concordant outcomes between the two arms (e.g., due to an unmeasured but stable baseline health status). Conversely, $\gamma_{i,j,x} < 1$ indicates negative residual association. In practice, $\gamma_{i,j,x}$ is treated as a tunable sensitivity parameter rather than estimated from the data; a common choice is to set $\gamma_{i,j,x}$ to a single value $\gamma$ across all $(i, j, x)$ and to vary $\gamma$ over a clinically plausible range to examine how sensitive the estimated Causal Gen Odds is to departures from conditional independence.

For applied analyses, we recommend reporting the estimate under conditional independence ($\gamma = 1$) alongside a robustness profile across a clinically motivated range of $\gamma$, and stating whether the qualitative conclusion holds throughout that range. Since $\gamma_{i,j,x}$ is not identifiable from trial data, its plausible range must be informed by substantive knowledge about the strength of unmeasured prognostic factors. Sources of such knowledge may include within-patient stability of the endpoint over time, test--retest or crossover data from similar populations, and the proportion of prognostic variation already captured by measured covariates. When shared unmeasured prognostic factors are expected to induce positive residual association between the potential outcomes, examining values of $\gamma$ greater than 1 may provide a particularly relevant sensitivity analysis.

Given marginal conditional probabilities $\Pr(Y_t < i \mid x)$ and $\Pr(Y_c < j \mid x)$ together with a value of $\gamma_{i,j,x}$, the bivariate Plackett formula recovers the joint probability \cite{plackett1965class}:
\begin{align}
	\Pr(Y_t < i,\, Y_c < j \mid x) =
	\begin{cases}
		\Pr(Y_t < i \mid x)\, \Pr(Y_c < j \mid x),
		& \gamma_{i,j,x} = 1, \\[8pt]
		\dfrac{-A(i,j,x) + \sqrt{A(i,j,x)^2 + B(i,j,x)}}
		      {2\,(1 - \gamma_{i,j,x})},
		& \gamma_{i,j,x} \neq 1,
	\end{cases}
	\label{eq5}
\end{align}
where:
\begin{align*}
	A(i,j,x) &= 1 + \bigl[\Pr(Y_t < i \mid x) +
	            \Pr(Y_c < j \mid x)\bigr]\,(\gamma_{i,j,x} - 1), \\
	B(i,j,x) &= 4\,(1 - \gamma_{i,j,x})\,\gamma_{i,j,x}\,
	            \Pr(Y_t < i \mid x)\,\Pr(Y_c < j \mid x).
\end{align*}

Combining Equations~\eqref{eq:identification} and~\eqref{eq5} yields the identification result (detailed proof in Appendix~\ref{app: A.1}):

\[
  \Pr(Y_t < Y_c)
  = \mathbb{E}_X\!\Bigl[\sum_{i=2}^{J}
        \Delta_{Y_{t}<Y_{c}}(i,\,X)\Bigr]
  = \int \sum_{i=2}^{J}
        \Delta_{Y_{t}<Y_{c}}(i,\,x)\,f_X(x)\,dx,
\]
where, for an ordinal outcome with $J$ categories ($1, 2, \ldots, J$), the index $i$ ranges from $2$ to $J$; the boundary term $i = 1$ vanishes since $\Pr(Y < 1 \mid \cdot) \equiv 0$. The function $\Delta_{Y_t < Y_c}$ is defined by
\[\Delta_{Y_{t}<Y_{c}}(i,x) =
\begin{cases}
	\Pr(Y_t < i \mid x)\Pr(Y_c < i+1 \mid x) - \\\ \ \ \ \Pr(Y_t < i \mid x)\Pr(Y_c < i \mid x), 
	& \gamma_{i,i+1,x}=1,\ \gamma_{i,i,x}=1 \\[8pt]
	\Pr(Y_t < i \mid x)\Pr(Y_c < i+1 \mid x) - \\\ \ \ \ \dfrac{-A(i,i,x) + \sqrt{A(i,i,x)^2 + B(i,i,x)}}{2(1-\gamma_{i,i,x})}, 
	& \gamma_{i,i+1,x}=1,\ \gamma_{i,i,x}\neq 1 \\[8pt]
	\dfrac{-A(i,i+1,x) + \sqrt{A(i,i+1,x)^2 + B(i,i+1,x)}}{2(1-\gamma_{i,i+1,x})} - \\\ \ \ \ \Pr(Y_t < i \mid x)\Pr(Y_c < i \mid x), 
	& \gamma_{i,i+1,x}\neq 1,\ \gamma_{i,i,x}=1 \\[8pt]
	\dfrac{-A(i,i+1,x) + \sqrt{A(i,i+1,x)^2 + B(i,i+1,x)}}{2(1-\gamma_{i,i+1,x})} - \\\ \ \ \ \dfrac{-A(i,i,x) + \sqrt{A(i,i,x)^2 + B(i,i,x)}}{2(1-\gamma_{i,i,x})}, 
	& \gamma_{i,i+1,x}\neq 1,\ \gamma_{i,i,x}\neq1
\end{cases}
\]

Similarly, the identification results for $\Pr(Y_{t} > Y_{c})$ and $\Pr(Y_{t} = Y_{c})$ follow by the same argument; the corresponding $\Delta$ functions and their derivations are given in Appendices~\ref{app: A.2} and~\ref{app: A.3}.

Following the plug-in estimation strategy of Section~2.3, the covariate expectation is replaced by the empirical distribution and the conditional probabilities by their model-based estimates, yielding
\[
  \widehat{\Pr}(Y_t < Y_c)
  = \frac{1}{n}\sum_{l=1}^{n}\sum_{i=2}^{J}
        \widehat{\Delta}_{Y_{t}<Y_{c}}(i,\,x_l),
\]
where $\widehat{\Delta}$ denotes $\Delta$ evaluated at the fitted conditional probabilities $\widehat{\Pr}(Y\leq j\mid T,X)$ from Equation~\eqref{eq:model}. The estimators $\widehat{\Pr}(Y_t > Y_c)$ and $\widehat{\Pr}(Y_t = Y_c)$ are obtained analogously. In implementation, each $\widehat{\Delta}(i, x_l)$ is evaluated using the subject-specific sensitivity parameters $\gamma_{i,j,x_l}$ at the threshold pairs required by the corresponding $\Delta$ function (e.g., $j=i$ and $j=i+1$ for $\Delta_{Y_t<Y_c}$); the constant-$\gamma$ case is recovered by setting $\gamma_{i,j,x_l} = \gamma$ uniformly, while heterogeneous values across subjects are accommodated by the same machinery when prior knowledge suggests $\gamma_{i,j,x_l}$ varies with $X$.

The estimated Causal Gen Odds is then:
\begin{eqnarray}
	\widehat{\text{Gen Odds}} 
	&=& \frac{\widehat{\Pr}(Y_{t} > Y_{c}) 
	          + \widehat{\Pr}(Y_{t} = Y_{c})/2}
	         {\widehat{\Pr}(Y_{t} < Y_{c}) 
	          + \widehat{\Pr}(Y_{t} = Y_{c})/2} \notag \\[6pt]
	&=& \frac{\dfrac{1}{n}\displaystyle\sum_{l=1}^{n}\sum_{i=2}^{J}
	          \widehat{\Delta}_{Y_t>Y_c}(i,x_l)
	          + \dfrac{1}{2n}\displaystyle\sum_{l=1}^{n}\sum_{i=1}^{J}
	          \widehat{\Delta}_{Y_t=Y_c}(i,x_l)}
	         {\dfrac{1}{n}\displaystyle\sum_{l=1}^{n}\sum_{i=2}^{J}
	          \widehat{\Delta}_{Y_t<Y_c}(i,x_l)
	          + \dfrac{1}{2n}\displaystyle\sum_{l=1}^{n}\sum_{i=1}^{J}
	          \widehat{\Delta}_{Y_t=Y_c}(i,x_l)}.\notag
\end{eqnarray}

At $\gamma = 1$ (i.e., $\gamma_{i,j,x} = 1$ for all $(i,j,x)$), the above estimator coincides with the conditional independence estimator of Section~2.3; for $\gamma \neq 1$, it reflects the impact of residual dependence on the Causal Gen Odds.

\section{Simulation}

\subsection{Simulation Setup}

We conducted a simulation study to evaluate the proposed estimators under two scenarios that share an identical data-generating mechanism but differ in the estimation strategy: (i) Scenario 1 implements the conditional-independence estimator described in Section~2.3, setting $\gamma_{i,j,x}=1$ across various data-generating settings; and (ii) Scenario 2 implements the sensitivity-analysis estimator described in Section~2.4, allowing for dependence between the potential outcomes by incorporating the data-generating value of $\gamma_{i,j,x}$.

\textbf{Population and covariates.} Each simulated dataset consisted of $n = 1000$ subjects, with 500 assigned to treatment and 500 to control. Two baseline covariates were drawn independently: a continuous covariate $X_1 \sim \mathrm{Uniform}(18, 80)$ and a binary covariate $X_2 \sim \mathrm{Bernoulli}(0.5)$, and treatment was randomly assigned in a balanced fashion.

\textbf{Marginal models for the potential outcomes.} For each subject, the conditional marginal distributions of the potential outcomes $Y_t, Y_c \in \{1, 2, 3\}$ (representing \textit{unchanged}, \textit{improved}, and \textit{cured}, respectively) were specified through a nonproportional cumulative logit model in which all coefficients vary across the two cut-points:
\[
  \logit\bigl\{\Pr(Y \leq j \mid T, X)\bigr\}
  = \alpha_j - \beta_{T,j}\, T - \beta_{1,j} X_1 - \beta_{2,j}X_2,
  \quad j = 1, 2,
\]
with $\alpha_1 = -0.5$, $\alpha_2 = 1.5$, $(\beta_{T,1}, \beta_{T,2}) = (2.0, 1.2)$, $(\beta_{1,1}, \beta_{1,2}) = (0.020, 0.025)$, and $(\beta_{2,1}, \beta_{2,2}) = (0.25, 0.35)$. Under this setup the proportional odds assumption is severely violated, so a single common odds ratio cannot summarize the treatment effect across the response distribution, motivating a global summary such as the Gen Odds. 

\textbf{Joint distribution and sensitivity parameter.} Given the marginals above and specified values of $\gamma_{i,j,x}$, the joint conditional distribution of $(Y_t, Y_c) \mid X$ was constructed through Equation~\eqref{eq5}. The pair $(Y_t, Y_c)$ was then drawn from the resulting $3 \times 3$ table, and the observed outcome was set to $Y_t$ for treated subjects and $Y_c$ for controls. We considered 11 settings for the true $\gamma_{i,j,x}$: nine fixed values applied uniformly across all $(i, j, x)$, $\gamma \in \{0.2, 0.4, 0.6, 0.8, 1.0, 2.0, 3.0, 4.0, 5.0\}$; and two settings in which each subject's $\gamma_{i,j,x_l}$ was drawn independently at random from either $\{0.2, 0.4, 0.6, 0.8\}$ (random low) or $\{2, 3, 4, 5\}$ (random high). Both the simulated data and the corresponding true Gen Odds (computed by Monte Carlo) drew $\gamma$ from the same distribution within each setting, so the reported truth represents the expected Gen Odds over that distribution.

\textbf{True value.} For each $\gamma_{i,j,x}$ setting, the true Causal Gen Odds was computed by Monte Carlo, using $10^8$ independent covariate draws to integrate the conditional probabilities. By the law of large numbers, the Monte Carlo error in the resulting truth is negligible relative to the simulation variability across replicates, and thus does not materially influence the reported bias, MSE, or coverage.

\textbf{Estimation strategies.} For each data-generating $\gamma$ setting, two estimation strategies were evaluated. First, the conditional-independence estimator described in Section~2.3 was implemented by setting $\gamma_{i,j,x_l}=1$ for every subject across all data-generating settings. This evaluates the performance of the conditional-independence estimator under varying degrees of dependence between $Y_t$ and $Y_c$. Second, the sensitivity-analysis estimator described in Section~2.4 was implemented by setting $\gamma_{i,j,x_l}$ to the value used in data generation. This evaluates the performance of the sensitivity-analysis estimator when the specified sensitivity parameter reflects the underlying dependence between the potential outcomes.

In practice, the true $\gamma$ is unknown, so analysts must scan a plausible range. Scenario~1 evaluates the conditional-independence estimator across varying degrees of dependence between the potential outcomes, while Scenario~2 evaluates the sensitivity-analysis estimator when the data-generating value of $\gamma$ is incorporated; together, they illustrate the role of the sensitivity parameter in the proposed framework.

\textbf{Replication and inference.} Each of the 22 conditions (11 true-$\gamma$ values $\times$ 2 estimation strategies) was replicated 1000 times. For each replicate, a 95\% confidence interval was constructed via 1000 nonparametric bootstrap resamples. Performance was summarized by bias (mean estimate minus the true Gen Odds for that condition), mean squared error (MSE), and empirical CI coverage. To guarantee that the two strategies produce identical numerical results when $\gamma_{\text{true}} = 1$, the data-generation seed was shared across strategies within each replicate.
\FloatBarrier

\subsection{Performance of the Conditional-Independence Estimator}

\begin{figure}[H]
    \centering
    \includegraphics[width=0.9\linewidth]{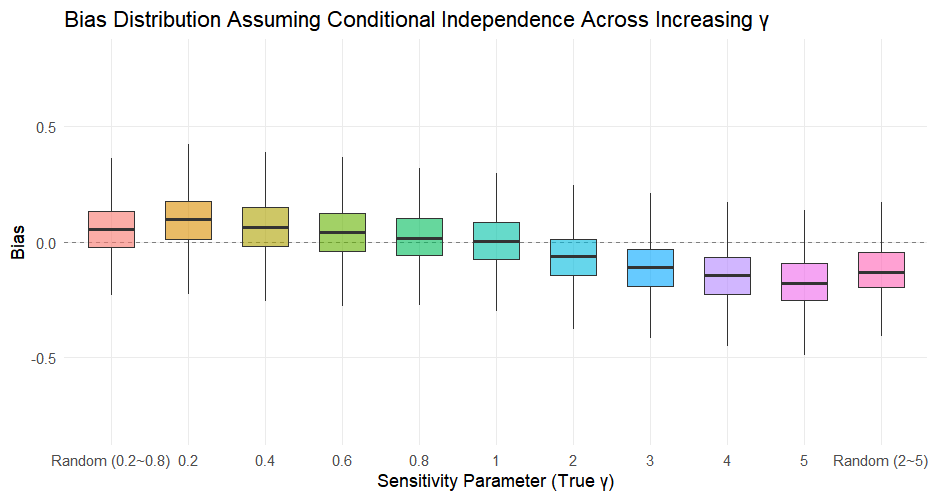}
    \caption{Distribution of estimated Gen Odds under the conditional independence assumption ($\gamma_{i,j,x_l} = 1$) across increasing values of the true $\gamma_{i,j,x}$. Each box summarizes 1000 simulated point estimates; the horizontal reference line indicates the true Gen Odds for the corresponding condition.}
    \label{fig:1}
\end{figure}

\autoref{fig:1} and \autoref{tab:appendix2} report the performance of the estimator when $\gamma_{i,j,x_l}$ is fixed at 1 for every subject, regardless of the true generating value. When the assumption actually holds ($\gamma_{\text{true}} = 1$), the estimator is approximately unbiased (bias $< 0.01$) with low MSE (0.01) and CI coverage close to the nominal level (95.5\%).

As $\gamma_{\text{true}}$ moves further from 1, performance deteriorates monotonically. For $\gamma_{\text{true}} > 1$ (positive residual association), the estimator underestimates the true Gen Odds, while for $\gamma_{\text{true}} < 1$ (negative association) it overestimates. At the extreme $\gamma_{\text{true}} = 5$, the bias reaches $-0.17$, MSE rises to 0.04, and coverage drops to 70.8\%. The random-$\gamma$ settings yield results consistent with averaging over the underlying $\gamma$ range, with coverage at 83.2\% (random high) and 92.8\% (random low). These findings demonstrate that routinely assuming conditional independence can introduce meaningful bias and undercoverage when residual dependence is nontrivial, and that the magnitude of this distortion is undetectable from the observed data alone.

\FloatBarrier

\subsection{Performance of the Sensitivity-Analysis Estimator}

\begin{figure}[H]
    \centering
    \includegraphics[width=0.9\linewidth]{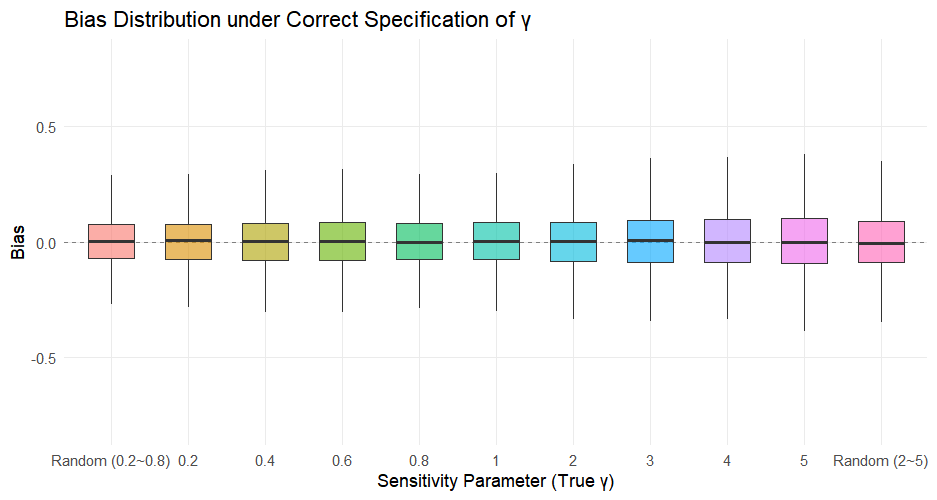}
    \caption{Distribution of estimated Gen Odds under the sensitivity-analysis estimator using the data-generating value of $\gamma_{i,j,x_l}$. Each box summarizes 1000 simulated point estimates; the horizontal reference line indicates the true Gen Odds for the corresponding condition.}
    \label{fig:2}
\end{figure}

\autoref{fig:2} and \autoref{tab:appendix3} present the corresponding results for Scenario~2, where the sensitivity-analysis estimator incorporates the data-generating value of $\gamma_{i,j,x_l}$. Across all 11 conditions, the bias is negligible ($< 0.01$), MSE remains in the range 0.01--0.02, and CI coverage stays close to the nominal 95\% level (93.7\%--96.1\%). Notably, the random-$\gamma$ settings---where each subject is assigned a different $\gamma_{i,j,x_l}$---are estimated with the same accuracy as the fixed-$\gamma$ conditions, demonstrating that the plug-in estimator readily accommodates subject-level heterogeneity in the sensitivity parameter without modification.

Taken together, the two parts show that the sensitivity-analysis estimator achieves its expected operating characteristics when the data-generating value of $\gamma$ is incorporated, whereas the conditional-independence estimator can exhibit increasing bias and undercoverage as residual dependence departs from $\gamma=1$. The practical implication is that reporting Gen Odds estimates across a range of plausible $\gamma$ values---as recommended in Section~2.4---is an essential safeguard against undetected residual dependence.

\section{Application}

We applied our Causal Gen Odds and previous ``Tournament Methods'' to data from the Action to Control Cardiovascular Risk in Diabetes (ACCORD) study. ACCORD was a multicenter randomized clinical trial designed to evaluate whether intensive glycemic control, lipid management, and intensive blood-pressure control could reduce cardiovascular events in individuals with type 2 diabetes at high cardiovascular risk \cite{accord2007design}. In the present analysis, we examined the effect of intensive versus standard glycemic control on one-year pre-tibial edema. The outcome was pre-tibial edema grade for the left and right feet, coded as a three-category ordinal outcome (1 = none, 2 = 1+, 3 = 2+ or higher; higher scores indicate more severe edema). Covariates were adjusted as in the ACCORD Study \cite{accord2007design}, including trial assignment (blood-pressure or lipid), intervention arm within each trial, clinical center network, history of cardiovascular events, and age as a continuous covariate. The conditional probabilities $\Pr(Y \le j \mid T, X)$ were estimated by fitting the cumulative logit model of Equation~\eqref{eq:model}.

The results for right foot pre-tibial edema are shown in \autoref{tab:10} (results for the left foot are in \autoref{tab:appendix4}). Our method, conditioning on the observed covariates, yields a Gen Odds of 0.909 (95\% CI: 0.880 to 0.940), consistent with the WMW Gen Odds of 0.908 (95\% CI: 0.878 to 0.938). All methods indicate a shift toward less severe edema under intensive treatment relative to standard treatment. Moreover, varying the sensitivity parameter $\gamma$ produces stable estimates, indicating robustness to moderate violations of conditional independence.

To further illustrate how the covariate-conditional and marginal comparisons can diverge, we conducted an additional hypothetical analysis based on the ACCORD data. We constructed a binary variable $X$ to induce a strong difference in the association between $X$ and the observed outcome $Y$ across treatment groups. In the control group, $X$ was strongly associated with higher outcome categories (e.g., when $Y = 2$ or $Y = 3$, $X = 1$ with 95\% probability), whereas in the treatment group, this relationship was reversed for $Y = 2$ ($X = 1$ with only 5\% probability). This deliberately constructed scenario was used solely as an illustration of how conditioning on a strongly prognostic variable can produce a conditional comparison that differs substantially from the marginal comparison; it should not be interpreted as an additional causal analysis of the ACCORD trial.

After conditioning on this hypothetical $X$, the estimated Gen Odds decreased from approximately 0.9 to 0.55 for both feet (see \autoref{tab:appendix5}). By contrast, the conventional Tournament Methods remained unchanged because they depend only on the marginal outcome distributions and do not condition on $X$. This illustration demonstrates that covariate-conditional and marginal Gen Odds can differ substantially when outcome distributions vary strongly across covariate strata.

\begin{table*}[!htbp]
	\centering
	\caption{Application of the Causal Gen Odds to right-foot pre-tibial edema in the ACCORD study}
	\label{tab:10}
	\begin{tabularx}{\textwidth}{XXXX}
		\toprule
		\begin{minipage}[t]{1.3in} \textbf{Method}\end{minipage}
		&
		&\begin{minipage}[t]{1.3in} \textbf{Measure of effect}\end{minipage}
		& \textbf{Estimates (95\%CI)}
		\\
		\midrule
		\begin{minipage}[t]{1.2in}\textbf{Causal Gen Odds}\end{minipage}
		&\begin{minipage}[t]{1.3in}\textbf{Conditional independence}\end{minipage}
		& & \\
		&$\gamma_{i, j, x} = 1$ &Causal Gen Odds &0.909 (0.880\ to\ 0.940) \\
		&&& \\
		&\begin{minipage}[t]{1.3in}\textbf{Sensitivity analysis}\end{minipage}
		& & \\
		&$\gamma_{i, j, x} = 2$ &Causal Gen Odds &0.908 (0.878\ to\ 0.940) \\
		&$\gamma_{i, j, x} = 3$ &Causal Gen Odds &0.907 (0.876\ to\ 0.939) \\
		&$\gamma_{i, j, x} = 4$ &Causal Gen Odds &0.905 (0.875\ to\ 0.939) \\
		&$\gamma_{i, j, x} = 5$ &Causal Gen Odds &0.904 (0.873\ to\ 0.938) \\
		&Random $\gamma_{i, j, x}$ &Causal Gen Odds &0.906 (0.875\ to\ 0.939) \\
		& & & \\
		\begin{minipage}[t]{1.3in}\textbf{Proportional odds ratios}\end{minipage}
		& &Common odds ratio &0.748 (0.677\ to\ 0.827) \\
		& & & \\
		\begin{minipage}[t]{1.3in}\textbf{Conventional Gen Odds}\end{minipage}
		& &Agresti Gen Odds &0.757 (0.689\ to\ 0.833) \\
		& &WMW Gen Odds &0.908 (0.878\ to\ 0.938) \\
		& & & \\
		\begin{minipage}[t]{1.3in}\textbf{Wilcoxon Mann-Whitney U test}\end{minipage}
		& &U Statistic &0.476 \\
		& & & \\
		\begin{minipage}[t]{1.3in}\textbf{Win ratio \& Win difference}\end{minipage}
		& &Win ratio &0.757\ (0.686\ to\ 0.839) \\
		& &Win difference &-0.048\ (-0.066\ to\ -0.031) \\
		& & & \\
		\begin{minipage}[t]{1.2in}\textbf{Win probability}\end{minipage}
		& &WinP(treatment) &0.476\ (0.467\ to\ 0.484) \\
		& &WinP(control) &0.524\ (0.516\ to\ 0.533) \\
		\bottomrule
	\end{tabularx}%
\end{table*}%

\section{Discussion}

This study defines the Causal Gen Odds as an individual-level estimand that complements the conventional population-level Gen Odds within the potential outcomes framework \cite{rubin1974estimating,holland1986statistics,hernan2010causal}. We establish its identification under conditional independence of the potential outcomes given baseline covariates and develop a sensitivity analysis that allows residual dependence to be characterized by an odds ratio $\gamma_{i,j,x}$ \cite{shen2013treatment,williamson1995analyzing,plackett1965class}. The simulations demonstrate the performance of the conditional-independence and sensitivity-analysis estimators across varying dependence structures. In the ACCORD Study, the Causal and conventional Gen Odds were nearly identical and remained stable across the examined sensitivity range, whereas the hypothetical-covariate analysis illustrates that covariate-conditional and marginal Gen Odds can differ substantially when outcome distributions vary strongly across covariate strata. Together, these results highlight the distinction between population-level marginal comparisons and individual-level treatment contrasts, while providing a framework for assessing the robustness of the latter to assumptions about the dependence between potential outcomes.

Our framework also differs fundamentally from existing covariate-adjustment strategies for tournament statistics. While stratified win ratios and inverse-probability-weighted win statistics use covariates to balance the two arms, they continue to contrast marginal outcome distributions \cite{dong2018stratified,gasparyan2021adjusted,wang2025adjusted}. In contrast, our approach uses covariates together with assumptions about the dependence between potential outcomes to identify the joint distribution of $(Y_t, Y_c)$, the fundamental object required when the target of inference is an individual-level contrast \cite{shen2013treatment,zhang2013assessing}. Because the win ratio, win difference, and win probability are all functionals of the same joint probabilities $q_{ij}$, the identification and sensitivity-analysis machinery developed here extends to these estimands with only notational changes, suggesting a pathway toward causal versions of the broader tournament family.

Several limitations should be noted. First, the conditional independence assumption between potential outcomes cannot be verified from observed data because $(Y_t,Y_c)$ are never jointly observed for the same individual. Although our framework incorporates a sensitivity analysis for departures from this assumption, the sensitivity parameter $\gamma_{i,j,x}$ is likewise not identifiable from the observed data. Consequently, the validity of causal conclusions depends on the plausibility of the assumed range of $\gamma$. Structured models where $\gamma$ varies systematically with covariates or cut-points based on clinical consensus represent a natural direction for future work. Second, when $X$ includes continuous covariates, estimation of the conditional outcome probabilities relies on the cumulative logit model in Equation~\eqref{eq:model}, and modeling continuous variables (such as age or blood pressure) with the wrong mathematical form may bias our estimator; incorporating more flexible or doubly robust estimation methods would strengthen this step. Finally, the framework is developed under randomization; extending it to observational studies would additionally require adequate control for confounding of treatment assignment—an assumption distinct from and additional to the conditional independence of potential outcomes considered here.

\section{Conclusions}

The Causal Gen Odds provides an individual-level estimand for ordinal outcomes that complements the population-level comparison summarized by the conventional Gen Odds. By combining covariate adjustment with explicit assumptions about the dependence between potential outcomes, the proposed framework enables identification and estimation of the Causal Gen Odds and provides a sensitivity analysis for assessing robustness to departures from conditional independence. This framework clarifies the distinction between marginal and individual-level treatment comparisons and provides a general approach for extending tournament-based measures toward individual-level causal inference.

\section*{Abbreviations}

ACCORD: Action to Control Cardiovascular Risk in Diabetes; 
BioLINCC: Biologic Specimen and Data Repository Information Coordinating Center;
CI: Confidence interval;
CMH: Cochran--Mantel--Haenszel;
EDSS: Expanded Disability Status Scale;
Gen Odds: Generalized Odds;
MSE: Mean squared error;
mRS: Modified Rankin Scale;
NHLBI: National Heart, Lung, and Blood Institute;
POLR: Proportional odds logistic regression;
RCT: Randomized controlled trial;
WMW: Wilcoxon--Mann--Whitney.

\section*{Declarations}

\subsection*{Trial registration}

The present study is a methodological study and did not involve the conduct of a clinical trial. The empirical application used data from the Action to Control Cardiovascular Risk in Diabetes (ACCORD) randomized controlled trial, registered at ClinicalTrials.gov (NCT00000620; registration date: October 27, 1999).

\subsection*{Ethics approval and consent to participate}

The original ACCORD study was approved by the institutional review boards or ethics committees at all participating clinical sites, and all participants provided written informed consent. The present study was a secondary analysis of de-identified ACCORD data obtained through the National Heart, Lung, and Blood Institute (NHLBI) BioLINCC repository.

\subsection*{Consent for publication}

Not applicable.

\subsection*{Availability of data and materials}

The ACCORD data analyzed in this study are available through the National Heart, Lung, and Blood Institute (NHLBI) Biologic Specimen and Data Repository Information Coordinating Center (BioLINCC). Access to the data requires submission and approval of a data request in accordance with BioLINCC requirements. The ACCORD study is available under BioLINCC accession number HLB01041317a. The simulation and analysis code used in this study are available from the corresponding authors upon reasonable request.

\subsection*{Competing interests}

The authors declare that they have no competing interests.

\subsection*{Funding}

This research received no specific grant from any funding agency in the public, commercial, or not-for-profit sectors.

\subsection*{Authors' contributions}

RH performed the methodological derivations, data processing, simulation studies, data analyses, and drafted and revised the manuscript. DS contributed to the conceptualization of the study and revision of the manuscript. LL and YY jointly supervised the study, contributed to its conceptualization, and critically revised the manuscript. LL and YY contributed equally as senior authors. All authors read and approved the final manuscript.

\subsection*{Acknowledgements}

Not applicable.

\bibliography{reference}

\appendix
\section{Calculation Details for the Illustrative Examples}
\label{appendix:calc}

\subsection*{Conventional and Causal Gen Odds for Cases 1 and 2}

\noindent\textbf{\autoref{tab:5}.}
For the conventional Gen Odds, the comparison probabilities are obtained from independent draws from the two marginal distributions and are therefore computed as products of the marginal probabilities. For instance, the marginal probabilities are $\Pr(Y_t=1)=0.39$, $\Pr(Y_t=2)=0.15$, $\Pr(Y_t=3)=0.46$, and $\Pr(Y_c=1)=0.24$, $\Pr(Y_c=2)=0.35$, $\Pr(Y_c=3)=0.41$. For example, $\Pr(\tilde{Y}_t=3,\tilde{Y}_c=1) =0.46\times0.24=0.1104$. Summing over wins, losses, and ties yields a conventional Gen Odds of $0.90$. 

Conversely, using the joint probabilities directly from \autoref{tab:5}:
\[
\Pr(Y_t>Y_c) = q_{21}+q_{31}+q_{32} = 0.0400+0.0800+0.2864 = 0.4064,
\]
\[
\Pr(Y_t<Y_c) = q_{12}+q_{13}+q_{23} = 0.0276+0.2424+0.0740 = 0.3440,
\]
\[
\Pr(Y_t=Y_c) = q_{11}+q_{22}+q_{33} = 0.1200+0.0360+0.0936 = 0.2496,
\]
\[
\text{Causal Gen Odds}=\frac{0.4064 + 0.5 \times 0.2496}{0.3440 + 0.5 \times 0.2496} = \frac{0.5312}{0.4688} = 1.13.
\]

\noindent\textbf{\autoref{tab:6}.}
Using independent draws from the two marginal distributions, the conventional Gen Odds is $1.12$. Using the joint distribution from \autoref{tab:6}:
\[
\Pr(Y_t>Y_c) = q_{21}+q_{31}+q_{32} = 0.0086+0.1230+0.0948 = 0.2264,
\]
\[
\Pr(Y_t<Y_c) = q_{12}+q_{13}+q_{23} = 0.2796+0.0120+0.0158 = 0.3074,
\]
\[
\Pr(Y_t=Y_c) = q_{11}+q_{22}+q_{33} = 0.0084+0.1656+0.2922 = 0.4662,
\]
\[
\text{Causal Gen Odds}=\frac{0.2264 + 0.5 \times 0.4662}{0.3074 + 0.5 \times 0.4662} = \frac{0.4595}{0.5405} = 0.85.
\]

\subsection*{Identification under Conditional Independence}

\noindent\textbf{Case 1 (\autoref{tab:5}).}
The true Causal Gen Odds is obtained by computing the conditional joint distribution within each stratum and averaging over $X$. Within stratum $X = x_1$ ($\Pr = 0.5$), summing the lower-left cells gives $\Pr(Y_t > Y_c \mid x_1) = 0.0640 + 0.0256 + 0.0016 = 0.0912$; the upper-right cells give $\Pr(Y_t < Y_c \mid x_1) = 0.0144 + 0.4752 + 0.1320 = 0.6216$; and the diagonal gives $\Pr(Y_t = Y_c \mid x_1) = 0.2304 + 0.0040 + 0.0528 = 0.2872$. Analogous calculations apply for $X = x_2$, where $\Pr(Y_t > Y_c \mid x_2) = 0.7216$, $\Pr(Y_t < Y_c \mid x_2) = 0.0664$, and $\Pr(Y_t = Y_c \mid x_2) = 0.2120$. Marginalization over $X$ yields the same population parameters ($\Pr(Y_t > Y_c) = 0.4064$, $\Pr(Y_t < Y_c) = 0.3440$, $\Pr(Y_t = Y_c) = 0.2496$), leading to a Causal Gen Odds of $1.13$. This matches the joint distribution in the ``Total'' panel while demonstrating that the total-level dependence is fully explained by the conditional strata. Using independent draws from the two marginal distributions in the
``Total'' panel yields the conventional Gen Odds of 0.90.

\noindent\textbf{Case 2 (\autoref{tab:6}).}
Following the same conditional stratification and marginalization procedure as described above for \autoref{tab:6}, the Causal Gen Odds equals 0.85, whereas the conventional Gen Odds based on independent draws from the two marginal distributions equals 1.12.

% -----------------------------------------------------------------------

\section{Mathematical Proofs}\label{sec:appendixa}

\setcounter{equation}{0}
\setcounter{table}{0}

\renewcommand\theequation{A-\arabic{equation}}
\renewcommand\thetable{A-\arabic{table}}

Let us define:
\begin{eqnarray}
	\Delta_{Y_{t}<Y_{c}}(i,x) &=& \Pr(Y_t < i, Y_c < i+1 \mid x) - \Pr(Y_t < i, Y_c < i \mid x),\notag \\
	\Delta_{Y_{t}>Y_{c}}(i,x) &=& \Pr(Y_t < i+1, Y_c < i \mid x) - \Pr(Y_t < i, Y_c < i \mid x),\notag \\
	\Delta_{Y_{t}=Y_{c}}(i,x) &=& \Pr(Y_t < i+1, Y_c < i+1 \mid x) - \Pr(Y_t < i, Y_c < i+1 \mid x) -\notag \\
	&& \Pr(Y_t < i+1, Y_c < i \mid x) + \Pr(Y_t < i, Y_c < i \mid x).\notag
\end{eqnarray}

\subsection{Derivation of \texorpdfstring{$\Pr(Y_t < Y_c)$}{Pr(Yt < Yc)}} \label{app: A.1}

For the general case (where $\gamma_{i, j, x}$ may take any positive value), we have:
\begin{eqnarray}
	\Pr(Y_{t} < Y_{c}) &=& \sum_{i < j} \Pr(Y_{t} = i, Y_{c} = j) \notag \\
    &=& \sum_{i=2}^{J} \mathbb{E}[\Pr(Y_{t} < i, Y_{c} = i \mid X)] \notag \\
	&=& \sum_{i=2}^{J} \int \Pr(Y_{t} < i, Y_{c} = i \mid X = x)\, f_X(x)\,dx \notag \\
	&=& \sum_{i=2}^{J} \int \big(\Pr(Y_{t} < i, Y_{c} < i + 1 \mid X = x) -\notag \\
	&& \Pr(Y_{t} < i, Y_{c} < i \mid X = x)\big)\, f_X(x)\,dx \notag \\
	&=&\sum_{i=2}^{J} \int \left(\frac{-A(i, i + 1,x) + \sqrt{A(i, i + 1,x)^2 + B(i, i + 1,x)}}{2 (1 - \gamma_{i, i + 1,x})} - \right.\notag\\
	&&\left.\frac{-A(i,i,x) + \sqrt{A(i,i,x)^2 + B(i,i,x)}}{2 (1 - \gamma_{i,i,x})}\right) f_X(x)\,dx. \notag
\end{eqnarray}

The above derivation uses the Plackett form of Equation~\eqref{eq5}; when $\gamma_{i,j,x} = 1$, the Plackett formula reduces to the product $\Pr(Y_t < i \mid x)\Pr(Y_c < j \mid x)$. Combining both cases, $\Pr(Y_t < Y_c)$ can be written as:
\[\Pr(Y_t < Y_c) = \int \sum_{i=2}^{J} \Delta_{Y_{t}<Y_{c}}(i,x)\, f_X(x)\,dx,\]

where:
\[\Delta_{Y_{t}<Y_{c}}(i,x) =
\begin{cases}
	\Pr(Y_t < i \mid x)\Pr(Y_c < i+1 \mid x) - \\\ \ \ \ \Pr(Y_t < i \mid x)\Pr(Y_c < i \mid x), 
	& \gamma_{i,i+1,x}=1,\ \gamma_{i,i,x}=1 \\[8pt]
	\Pr(Y_t < i \mid x)\Pr(Y_c < i+1 \mid x) - \\\ \ \ \ \dfrac{-A(i,i,x) + \sqrt{A(i,i,x)^2 + B(i,i,x)}}{2(1-\gamma_{i,i,x})}, 
	& \gamma_{i,i+1,x}=1,\ \gamma_{i,i,x}\neq 1 \\[8pt]
	\dfrac{-A(i,i+1,x) + \sqrt{A(i,i+1,x)^2 + B(i,i+1,x)}}{2(1-\gamma_{i,i+1,x})} - \\\ \ \ \ \Pr(Y_t < i \mid x)\Pr(Y_c < i \mid x), 
	& \gamma_{i,i+1,x}\neq 1,\ \gamma_{i,i,x}=1 \\[8pt]
	\dfrac{-A(i,i+1,x) + \sqrt{A(i,i+1,x)^2 + B(i,i+1,x)}}{2(1-\gamma_{i,i+1,x})} - \\\ \ \ \ \dfrac{-A(i,i,x) + \sqrt{A(i,i,x)^2 + B(i,i,x)}}{2(1-\gamma_{i,i,x})}, 
	& \gamma_{i,i+1,x}\neq 1,\ \gamma_{i,i,x}\neq1
\end{cases}
\]

\subsection{Derivation of \texorpdfstring{$\Pr(Y_t > Y_c)$}{Pr(Yt > Yc)}} \label{app: A.2}
Similarly, for the general case (where $\gamma_{i, j, x}$ may take any positive value), we have:
\begin{eqnarray}
	\Pr(Y_{t} > Y_{c}) &=& \sum_{i > j} \Pr(Y_{t} = i, Y_{c} = j) \notag \\
	&=& \sum_{i=2}^{J} \mathbb{E}[\Pr(Y_{t} = i, Y_{c} < i \mid X)] \notag \\
	&=& \sum_{i=2}^{J} \int \Pr(Y_{t} = i, Y_{c} < i \mid X = x)\, f_X(x)\,dx \notag \\
	&=& \sum_{i=2}^{J} \int \big(\Pr(Y_{t} < i + 1, Y_{c} < i \mid X = x) -\notag \\
	&& \Pr(Y_{t} < i, Y_{c} < i \mid X = x)\big)\, f_X(x)\,dx \notag \\
	&=&\sum_{i=2}^{J} \int \left(\frac{-A(i+1, i,x) + \sqrt{A(i+1, i,x)^2 + B(i+1, i,x)}}{2 (1 - \gamma_{i+1, i,x})} - \right.\notag\\
	&&\left.\frac{-A(i,i,x) + \sqrt{A(i,i,x)^2 + B(i,i,x)}}{2 (1 - \gamma_{i,i,x})}\right) f_X(x)\,dx. \notag
\end{eqnarray}

So:
\begin{eqnarray}
	\Pr(Y_{t} > Y_{c})&=&\int \sum_{i=2}^{J} \big[ \Pr(Y_t < i+1, Y_c < i \mid x) - \Pr(Y_t < i, Y_c < i \mid x) \big]\, f_X(x)\,dx \notag \\
	&=& \int \sum_{i=2}^{J} \Delta_{Y_{t}>Y_{c}}(i,x)\, f_X(x)\,dx,\notag
\end{eqnarray}

where:
\[\Delta_{Y_{t}>Y_{c}}(i,x) =
\begin{cases}
	\Pr(Y_t < i+1 \mid x)\Pr(Y_c < i \mid x) - \\\ \ \ \ \Pr(Y_t < i \mid x)\Pr(Y_c < i \mid x), 
	& \gamma_{i+1,i,x}=1,\ \gamma_{i,i,x}=1 \\[8pt]
	\Pr(Y_t < i+1 \mid x)\Pr(Y_c < i \mid x) - \\\ \ \ \ \dfrac{-A(i,i,x) + \sqrt{A(i,i,x)^2 + B(i,i,x)}}{2(1-\gamma_{i,i,x})}, 
	& \gamma_{i+1,i,x}=1,\ \gamma_{i,i,x}\neq 1 \\[8pt]
	\dfrac{-A(i+1,i,x) + \sqrt{A(i+1,i,x)^2 + B(i+1,i,x)}}{2(1-\gamma_{i+1,i,x})} - \\\ \ \ \ \Pr(Y_t < i \mid x)\Pr(Y_c < i \mid x), 
	& \gamma_{i+1,i,x}\neq 1,\ \gamma_{i,i,x}=1 \\[8pt]
	\dfrac{-A(i+1,i,x) + \sqrt{A(i+1,i,x)^2 + B(i+1,i,x)}}{2(1-\gamma_{i+1,i,x})} - \\\ \ \ \ \dfrac{-A(i,i,x) + \sqrt{A(i,i,x)^2 + B(i,i,x)}}{2(1-\gamma_{i,i,x})}, 
	& \gamma_{i+1,i,x}\neq 1,\ \gamma_{i,i,x}\neq1
\end{cases}
\]

\subsection{Derivation of \texorpdfstring{$\Pr(Y_t = Y_c)$}{Pr(Yt = Yc)}}\label{app: A.3}
Similarly, for the general case (where $\gamma_{i, j, x}$ may take any positive value), we have:
\begin{eqnarray}
	\Pr(Y_{t} = Y_{c}) &=& \sum_{i = j} \Pr(Y_{t} = i, Y_{c} = j) \notag \\
	&=& \sum_{i=1}^{J} \int \Pr(Y_{t} = i, Y_{c} = i \mid X = x)\, f_X(x)\,dx \notag \\
	&=& \sum_{i=1}^{J} \int \big(\Pr(Y_{t} < i + 1, Y_{c} < i + 1 \mid X = x) -\notag \\
	&& \Pr(Y_{t} < i + 1, Y_{c} < i \mid X = x) -\notag \\
	&& \Pr(Y_{t} < i, Y_{c} < i + 1 \mid X = x) + \notag \\
	&& \Pr(Y_{t} < i, Y_{c} < i \mid X = x)\big)\, f_X(x)\,dx \notag \\
	&=& \sum_{i=1}^{J} \int \left(\frac{-A(i+1, i + 1,x) + \sqrt{A(i+1, i + 1,x)^2 + B(i+1, i + 1,x)}}{2 (1 - \gamma_{i+1, i + 1,x})} - \right.\notag\\
	&&\frac{-A(i,i+1,x) + \sqrt{A(i,i+1,x)^2 + B(i,i+1,x)}}{2 (1 - \gamma_{i,i+1,x})} - \notag \\	
	&&\frac{-A(i+1, i,x) + \sqrt{A(i+1, i,x)^2 + B(i+1, i,x)}}{2 (1 - \gamma_{i+1, i,x})} + \notag\\
	&&\left.\frac{-A(i,i,x) + \sqrt{A(i,i,x)^2 + B(i,i,x)}}{2 (1 - \gamma_{i,i,x})}\right) f_X(x)\,dx. \notag
\end{eqnarray}

So:
\begin{eqnarray}
	\Pr(Y_{t} = Y_{c}) &=& \int \sum_{i=1}^{J} \Delta_{Y_{t}=Y_{c}}(i,x)\, f_X(x)\,dx. \notag
\end{eqnarray}

In practice, $\Pr(Y_{t} = Y_{c})$ can be computed more directly as:
\begin{eqnarray}
	\Pr(Y_{t} = Y_{c}) = 1 - \Pr(Y_{t} > Y_{c}) - \Pr(Y_{t} < Y_{c}).\notag
\end{eqnarray}

\section{Supplementary Tables}\label{sec:appendixb}

\begin{table*}[!htbp]
	\centering
	\caption{Performance of the Causal Gen Odds estimator under the conditional   independence assumption ($\gamma_{i,j,x_l} = 1$ for all subjects). Simulations use $n = 1000$ subjects per dataset (500 per arm), 1000 replicates, and 1000 bootstrap resamples for each 95\% CI.}
	\label{tab:appendix2}
	\begin{tabularx}{\textwidth}{lXcccc}
    \toprule
    Dependence & True $\gamma_{i,j,x}$ & True value & Bias & MSE & Coverage (\%) \\
    \midrule
    Independence
      & $\gamma_{i,j,x} = 1$            & 1.85 & 
      $<$ 0.01 & 0.01 & 95.5 \\
    \addlinespace
    Positive ($\gamma > 1$)
      & $\gamma_{i,j,x} = 2$            & 1.91 & $-0.06$ & 0.02 & 91.5 \\
      & $\gamma_{i,j,x} = 3$            & 1.96 & $-0.11$ & 0.03 & 86.0 \\
      & $\gamma_{i,j,x} = 4$            & 2.00 & $-0.14$ & 0.03 & 78.5 \\
      & $\gamma_{i,j,x} = 5$            & 2.02 & $-0.17$ & 0.04 & 70.8 \\
      & Random $\gamma \in \{2,3,4,5\}$ & 1.97 & $-0.12$ & 0.03 & 83.2 \\
    \addlinespace
    Negative ($\gamma < 1$)
      & $\gamma_{i,j,x} = 0.2$          & 1.76 & 0.10 & 0.02 & 85.6 \\
      & $\gamma_{i,j,x} = 0.4$          & 1.79 & 0.07 & 0.02 & 90.9 \\
      & $\gamma_{i,j,x} = 0.6$          & 1.81 & 0.04 & 0.02 & 93.8 \\
      & $\gamma_{i,j,x} = 0.8$          & 1.83 & 0.02 & 0.01 & 95.3 \\
      & Random $\gamma \in \{0.2,0.4,0.6,0.8\}$ & 1.79 & 0.06 & 0.02 & 92.8 \\
    \bottomrule
  \end{tabularx}
\end{table*}%

\begin{table*}[!htbp]
  \centering
  \caption{Performance of the Causal Gen Odds estimator using the data-generating value of gamma. Simulations use $n = 1000$
  subjects per dataset (500 per arm), 1000 replicates, and 1000 bootstrap
  resamples for each 95\% CI.}
  \label{tab:appendix3}
 \begin{tabularx}{\textwidth}{lXcccc}
    \toprule
    Dependence & True $\gamma_{i,j,x}$ & True value & Bias & MSE & Coverage (\%) \\
    \midrule
    Independence
      & $\gamma_{i,j,x} = 1$            & 1.85 & 
      $<$ 0.01 & 0.01 & 95.5 \\
    \addlinespace
    Positive ($\gamma > 1$)
      & $\gamma_{i,j,x} = 2$            & 1.91 & $<$ 0.01 & 0.02 & 94.9 \\
      & $\gamma_{i,j,x} = 3$            & 1.96 & $<$ 0.01 & 0.02 & 95.0 \\
      & $\gamma_{i,j,x} = 4$            & 2.00 & $<$ 0.01 & 0.02 & 94.9 \\
      & $\gamma_{i,j,x} = 5$            & 2.02 & $<$ 0.01 & 0.02 & 95.1 \\
      & Random $\gamma \in \{2,3,4,5\}$ & 1.97 & $<$ 0.01 & 0.02 & 93.7 \\
    \addlinespace
    Negative ($\gamma < 1$)
      & $\gamma_{i,j,x} = 0.2$          & 1.76 & $<$ 0.01 & 0.01 & 96.1 \\
      & $\gamma_{i,j,x} = 0.4$          & 1.79 & $<$ 0.01 & 0.01 & 95.8 \\
      & $\gamma_{i,j,x} = 0.6$          & 1.81 & $<$ 0.01 & 0.01 & 95.1 \\
      & $\gamma_{i,j,x} = 0.8$          & 1.83 &  
      $<$ 0.01 & 0.01 & 95.5 \\
      & Random $\gamma \in \{0.2,0.4,0.6,0.8\}$ & 1.79 & $<$ 0.01 & 0.01 & 94.6 \\
    \bottomrule
  \end{tabularx}
\end{table*}

\begin{table*}[!htbp]
	\centering
	\caption{Application of the Causal Gen Odds to left-foot pre-tibial edema in the ACCORD study}
	\label{tab:appendix4}
	\begin{tabularx}{\textwidth}{XXXX}
		\toprule
		\begin{minipage}[t]{1.3in} \textbf{Method}\end{minipage}
		& &\begin{minipage}[t]{1.3in} \textbf{Measure of effect}\end{minipage}
		& \textbf{Estimates (95\%CI)} \\
		\midrule
		\begin{minipage}[t]{1.2in}\textbf{Causal Gen Odds}\end{minipage}
		&\begin{minipage}[t]{1.3in}\textbf{Conditional independence}\end{minipage}
		& & \\
		&$\gamma_{i, j, x} = 1$ &Causal Gen Odds &0.919 (0.888\ to\ 0.948) \\
		&&& \\
		&\begin{minipage}[t]{1.3in}\textbf{Sensitivity analysis}\end{minipage}
		& & \\
		&$\gamma_{i, j, x} = 2$ &Causal Gen Odds &0.917 (0.886\ to\ 0.947) \\
		&$\gamma_{i, j, x} = 3$ &Causal Gen Odds &0.916 (0.884\ to\ 0.946) \\
		&$\gamma_{i, j, x} = 4$ &Causal Gen Odds &0.915 (0.882\ to\ 0.946) \\
		&$\gamma_{i, j, x} = 5$ &Causal Gen Odds &0.913 (0.881\ to\ 0.945) \\
		&Random $\gamma_{i, j, x}$ &Causal Gen Odds &0.915 (0.884\ to\ 0.946) \\
		& & & \\
		\begin{minipage}[t]{1.3in}\textbf{Proportional odds ratios}\end{minipage}
		& &Common odds ratio &0.775 (0.702\ to\ 0.856) \\
		& & & \\
		\begin{minipage}[t]{1.3in}\textbf{Conventional Gen Odds}\end{minipage}
		& &Agresti Gen Odds &0.785 (0.715\ to\ 0.862) \\
		& &WMW Gen Odds &0.918 (0.888\ to\ 0.949) \\
		& & & \\
		\begin{minipage}[t]{1.3in}\textbf{Wilcoxon Mann-Whitney U test}\end{minipage}
		& &U Statistic &0.479 \\
		& & & \\
		\begin{minipage}[t]{1.3in}\textbf{Win ratio \& Win difference}\end{minipage}
		& &Win ratio &0.785\ (0.711\ to\ 0.863) \\
		& &Win difference &-0.043\ (-0.060\ to\ -0.026) \\
		& & & \\
		\begin{minipage}[t]{1.2in}\textbf{Win probability}\end{minipage}
		& &WinP(treatment) &0.479\ (0.470\ to\ 0.486) \\
		& &WinP(control) &0.521\ (0.514\ to\ 0.530) \\
		\bottomrule
	\end{tabularx}%
\end{table*}%

\begin{table*}[!htbp]
	\centering
	\caption{Illustrative analysis of the ACCORD data using a hypothetical binary variable}
	\label{tab:appendix5}
	\begin{tabularx}{\textwidth}{XXXX}
		\toprule
		\begin{minipage}[t]{1.2in} \textbf{Foot}\end{minipage}
		& &\begin{minipage}[t]{1.2in} \textbf{Measure of effect}\end{minipage}
		& \textbf{Estimates (95\%CI)} \\
		\midrule
		\begin{minipage}[t]{1.2in}\textbf{Right foot pre-tibial edema}\end{minipage}
		&\begin{minipage}[t]{1.3in}\textbf{Conditional independence}\end{minipage}
		& & \\
		&$\gamma_{i, j, x} = 1$ &Causal Gen Odds &0.553 (0.524\ to\ 0.585) \\
		&&& \\
		&\begin{minipage}[t]{1.3in}\textbf{Sensitivity analysis}\end{minipage}
		& & \\
		&$\gamma_{i, j, x} = 2$ &Causal Gen Odds &0.548 (0.519\ to\ 0.582) \\
		&$\gamma_{i, j, x} = 3$ &Causal Gen Odds &0.544 (0.513\ to\ 0.579) \\
		&$\gamma_{i, j, x} = 4$ &Causal Gen Odds &0.540 (0.509\ to\ 0.576) \\
		&$\gamma_{i, j, x} = 5$ &Causal Gen Odds &0.537 (0.505\ to\ 0.573) \\
		&Random $\gamma_{i, j, x}$ &Causal Gen Odds &0.542 (0.512\ to\ 0.577) \\
		&& & \\
		\begin{minipage}[t]{1.2in}\textbf{Left foot pre-tibial edema}\end{minipage}
		&\begin{minipage}[t]{1.3in}\textbf{Conditional independence}\end{minipage}
		& & \\
		&$\gamma_{i, j, x} = 1$ &Causal Gen Odds &0.559 (0.532\ to\ 0.589) \\
		&&& \\
		&\begin{minipage}[t]{1.3in}\textbf{Sensitivity analysis}\end{minipage}
		& & \\
		&$\gamma_{i, j, x} = 2$ &Causal Gen Odds &0.556 (0.525\ to\ 0.585) \\
		&$\gamma_{i, j, x} = 3$ &Causal Gen Odds &0.549 (0.519\ to\ 0.581) \\
		&$\gamma_{i, j, x} = 4$ &Causal Gen Odds &0.544 (0.514\ to\ 0.578) \\
		&$\gamma_{i, j, x} = 5$ &Causal Gen Odds &0.541 (0.510\ to\ 0.575) \\
		&Random $\gamma_{i, j, x}$ &Causal Gen Odds &0.547 (0.517\ to\ 0.580) \\
		\bottomrule
	\end{tabularx}%
\end{table*}%

\end{document}